\documentclass[manuscript]{acmart}
\AtBeginDocument{%
  \providecommand\BibTeX{{%
    \normalfont B\kern-0.5em{\scshape i\kern-0.25em b}\kern-0.8em\TeX}}}

\setcopyright{acmcopyright}
\copyrightyear{2023}
\acmYear{2023}
\acmDOI{XXXXXXX.XXXXXXX}

\acmConference[DIS '23]{ACM Designing Interactive Systems}{10-14 July, 2023}{Pittsburgh, PA}
\acmBooktitle{DIS '23: ACM Designing Interactive Systems,
 {10-14 July, 2023 Pittsburgh, PA} 
\acmPrice{15.00}
\acmISBN{978-1-4503-XXXX-X/18/06}
}

\begin{document}

\title{Bridging Data and Experiences: Engaging Youth in Digital Civics through Participatory Mapmaking for Resilience} 


\author{Mohsin Yousufi}
\email{yousufi@gatech.edu}
\orcid{0000-0002-7738-4087}
\affiliation{%
  \institution{Georgia Institute of Technology}
  \city{Atlanta}
  \state{Georgia}
  \country{USA}
}

\author{Yanni Loukissas}
\email{yanni.loukissas@lmc.gatech.edu}
\affiliation{%
  \institution{Georgia Institute of Technology}
  \city{Atlanta}
  \state{Georgia}
  \country{USA}
}
\author{Allen Hyde}
\email{allen.hyde@hsoc.gatech.edu}
\affiliation{%
  \institution{Georgia Institute of Technology}
  \city{Atlanta}
  \state{Georgia}
  \country{USA}
}

\renewcommand{\shortauthors}{Yousufi et al.}

\begin{abstract}
The paper reports on a participatory map-making workshop with middle school students in Savannah, Georgia. The workshop, which took place over a seven week period as part of Youth Advocacy for Resilience to Disasters (YARD) program, students worked together to make maps that examine local natural disasters and their consequences. We developed a map-making platform that enabled the youth to leverage spatial data on local environmental, economic and social issues, as a means of interrogating their existing stories about disasters and their  impacts. This allowed the students to share their specific experiences of the city, while also inquiring about the validity of such experiences. Through the use of our platform, we illustrate the potential of maps as representations that bridge two worlds, that of data and experience. We argue that Map Spot leverages existing data to help youth surface and communicate what they already know, but were otherwise unable to articulate.

\end{abstract}



\maketitle

\section{Introduction}

Existing research shows that visualization tools can help youth understand the effects of natural and man-made disasters, the value of resilience, and potential for improvement strategies \cite{mcallister_data_2019}. Generally, these tools are presented as a dashboard of indicators or metrics that reflect community needs and priorities, but are aimed at individual adult decision makers \cite{mcallister_data_2019}. Our project introduces a new map-making data visualization platform, (removed: name of our mapmaking platform), that is both accessible to youth and explicitly collaborative. We believe that it can help youth, especially from Black, Indigenous and People of Color (BIPOC) communities, reflect on the effects of intersecting disasters in their communities and imagine what resilience to these disasters might mean for them. 

We are particularly interested in the role of the maps as a bridge between the civic data, such as census,  and the everyday lived experience of the youth. The collaborative nature of the tool allowed us to create an "open data setting": a space in which participants can draw together their personal experiences and data \cite{loukissas_open_2021}. In the following, we describe how in a workshop for our project the youth, (initially hesitant to share their experiences of the city), developed an engaged and lively discussion around disasters and resilience, while reflecting on their own experiences of the neighborhoods. We argue that the co-creation of maps played a key role in this transition, and can lead to meaningfully engage youth, and other social groups, in civic discourses contributing to resilient communities. 

\section{Related work}

Digital civics, an emerging research area that blends digital media and planning, has sought to create data-driven public tools and services, which encourage active participation in the co-production of knowledge for local governance \cite{vlachokyriakos_digital_2016, dickinson_inclusion_2018,puussaar_making_2018, gurstein_open_2011}. Interventions in this space have looked at using citizens for participatory sensing “specifically, making data meaningful and useful for non-expert”\cite{coulson_stop_2018} and including young people in design process \cite{dow_scaffolding_2022}.

Similarly, there is also work on using data visualizations as part of digital civics, where Huron et al. explored how to make data visualizations democratic and engaging to nonexperts \cite{huron_constructive_2014}. Some projects that have built on this work include, ManyEyes – users upload and share their viz and collaborate on this \cite{viegas_manyeyes_2007}, Narrative visualizations – using visualization to tell stories from non-experts \cite{segel_narrative_2010}) .

 \begin{figure}
    \includegraphics[width=0.65\textwidth]{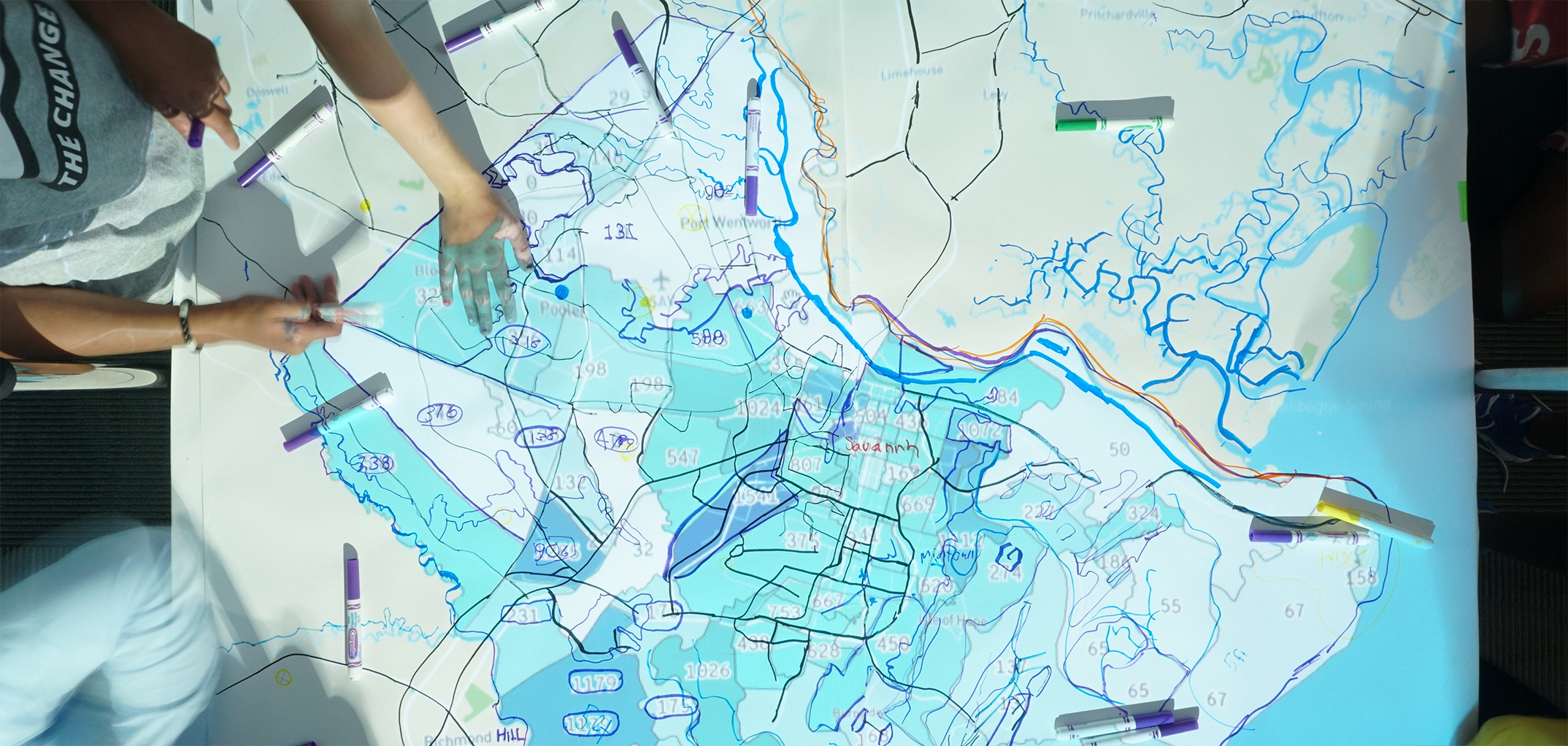}
    \caption{A view of the map-making platform with a "data layer" loaded}
  \label{fig:my_label}
\end{figure}

\section{The Workshop Session}

The project was part of the Youth Advocacy for Resilience to Disasters (YARDs) program, as part of NSF CIVCS challenge, conducted at a middle school in Savannah during Fall 2022. The program was led by a team from Georgia Tech, Savannah State University, local teachers and community partners, with a focus on engaging youth in disaster resilience planning through map-making and advocacy. This paper reports on one of the three map-making sessions out of the 14-session after-school program.

In this session the students mapped Savannah and discussed their stories using our map-making platform. We had pre-loaded 37 data sets visualized in forms of maps onto the platform. The data included, but not limited to: location of schools, community centers, hospital, pollution levels, cancer risk, demographic distribution, housing sizes, zoning. These data layers were limited to Chatham county, where all our participants lived. These sessions were led by the authors and the middle school teachers, and were recorded and documented. The names of participants are changed so as to preserve anonymity.  

\subsection{Background}

In a classroom typically used for art instruction, we conducted this 90 minute session. The classroom itself was only slightly rearranged to accommodate (the map-making platform). This platform is comprised of 3 important parts, 1) The projector 2) The drawing surface 3) Maps and Visualizations. We use a short-throw project mounted vertically on TV stand to project the image on a flat surface, in this case two tables covered with paper. For the projector image to be clear, the room must have limited lighting. Once set up, participants can use a range of media to draw, sketch, doodle and write on the paper and explore data visualizations about the places over layed over their drawings. In this session, we set up one (the map-making platform) station at the far end of the room.

\section{Discussion} 
Our project provides initial insights into  how collaborative mapping can contribute to digital civics work related to youth's knowledge of spaces and experiences. Our interests are two-fold: 1) Inquiring into the role of youth in digital civics, by confronting what they can contribute, how they might benefit from making a contribution, as well as the challenges and limitations they face,  2) Understanding Map Spot as a mechanism for surfacing, communicating experiences, connecting them to data and leveraging them in place-based claims.

\subsection{\textit{Role of Youth}}

We found that the young people we worked with were initially hesitant but. when engaged further in one-on-one interactions, did have experience and stories to share. The youth are politically and culturally aware of the histories of the places they live in \cite{kambunga_participatory_2020}. Moreover, we found that when they were allowed to guide their own explorations of data about the places they live, they were able to identify and better understand the issues that affect them and those they care about—in a way that they were previously not able to do (when asked about these issues abstractly). For instance the participants had initially limited reaction to pollution layer but when they identified the people and communities that lived there, their involvement to the data changed; for example Trisha (a student) who was initially quiet, reacted with an exclamation "See!" and hitting the desk with her hand she identified the trend of predominantly-Black neighborhoods having the worst air pollution, income, cancer risk and property distribution.  

We also find the initial hesitance of the youth and the strategies to work through them is a possible area of inquiry for researchers and designers attempting to engage in civics work with youth. We offer one reason why this might be so. For some of the young people we worked with, this was the first time that they were listened to in a serious way. There is a history of youth being involved in civics projects only for organisations and groups afterwards to disregard their voices \cite{dow_scaffolding_2022, kambunga_participatory_2020}. This also sometimes extends to their family lives; A participant later revealed they wouldn't discuss civic issues learned in the project at home because their mother would scold them saying "I don't have time for this."

\subsection{\textit{Dialog of data and experiences}}

Our project also illustrates how collaborative mapping is an effective tool to engage citizens with the issues and data about the places they live in. Community members have limited resources and time in day to day lives to engage in precises data collections methods or interact with complicated data sets \cite{coulson_stop_2018}. By putting data in context Map Spot allows peoples (especially groups such as youth or non-experts) to have informed dialogues. This also enabled participants to engage with data, as well as find areas of collective concern that might be unexpected by facilitators. We had anticipated that the data on flood levels would be the focus of the conversation for the group, but instead the students found data on pollution to be more relevant, which was a phenomenon that at least most of the group had a direct experience with. As researchers it was important for us to gauge how specific social groups had different concerns that would engage them and move them to collective action \cite{balestrini_city_2017-1}. The discovery of what was relevant to the youth in this specific setting emerged only through putting their experiences in dialog with the data. [EXPLAIN FURTHER!]

Through the collaborative and discursive process of creating maps using our platform, participants, researchers, and teachers were able to arrive at a shared concern. This concern was informed by both, the experiences of the students and the data available on the platform. 

\section{Conclusion}

This project sheds light on the potential of Map Spot as a tool for engaging youth in digital civics work and for making data accessible to non-expert community members. Through the collaborative process of creating maps with an interactive data platform, youth were able to identify and better understand the issues that affect them and those they care about. The project also highlights the importance of listening to the voices of youth and providing them with a platform to engage in meaningful dialogue about civic issues. The role of youth is key to building resilient communities, especially marginalized communities, and digital civics provides a particularly unique opportunity for greater civic participation by the youth. The use of maps as a visualization tool proved to be an effective way to engage citizens with the data about the places they live in, and to identify areas of collective concern.


\bibliographystyle{ACM-Reference-Format}
\bibliography{references_dis_wip}


\begin{thebibliography}{13}


\ifx \showCODEN    \undefined \def \showCODEN     #1{\unskip}     \fi
\ifx \showDOI      \undefined \def \showDOI       #1{#1}\fi
\ifx \showISBNx    \undefined \def \showISBNx     #1{\unskip}     \fi
\ifx \showISBNxiii \undefined \def \showISBNxiii  #1{\unskip}     \fi
\ifx \showISSN     \undefined \def \showISSN      #1{\unskip}     \fi
\ifx \showLCCN     \undefined \def \showLCCN      #1{\unskip}     \fi
\ifx \shownote     \undefined \def \shownote      #1{#1}          \fi
\ifx \showarticletitle \undefined \def \showarticletitle #1{#1}   \fi
\ifx \showURL      \undefined \def \showURL       {\relax}        \fi
\providecommand\bibfield[2]{#2}
\providecommand\bibinfo[2]{#2}
\providecommand\natexlab[1]{#1}
\providecommand\showeprint[2][]{arXiv:#2}

\bibitem[Balestrini et~al\mbox{.}(2017)]%
        {balestrini_city_2017-1}
\bibfield{author}{\bibinfo{person}{Mara Balestrini}, \bibinfo{person}{Yvonne
  Rogers}, \bibinfo{person}{Carolyn Hassan}, \bibinfo{person}{Javi Creus},
  \bibinfo{person}{Martha King}, {and} \bibinfo{person}{Paul Marshall}.}
  \bibinfo{year}{2017}\natexlab{}.
\newblock \showarticletitle{A {City} in {Common}: {A} {Framework} to
  {Orchestrate} {Large}-scale {Citizen} {Engagement} around {Urban} {Issues}}.
  In \bibinfo{booktitle}{\emph{Proceedings of the 2017 {CHI} {Conference} on
  {Human} {Factors} in {Computing} {Systems}}} \emph{(\bibinfo{series}{{CHI}
  '17})}. \bibinfo{publisher}{Association for Computing Machinery},
  \bibinfo{address}{New York, NY, USA}, \bibinfo{pages}{2282--2294}.
\newblock
\showISBNx{978-1-4503-4655-9}
\urldef\tempurl%
\url{https://doi.org/10.1145/3025453.3025915}
\showDOI{\tempurl}


\bibitem[Coulson et~al\mbox{.}(2018)]%
        {coulson_stop_2018}
\bibfield{author}{\bibinfo{person}{Saskia Coulson}, \bibinfo{person}{Mel
  Woods}, \bibinfo{person}{Michelle Scott}, \bibinfo{person}{Drew Hemment},
  {and} \bibinfo{person}{Mara Balestrini}.} \bibinfo{year}{2018}\natexlab{}.
\newblock \showarticletitle{Stop the {Noise}! {Enhancing} {Meaningfulness} in
  {Participatory} {Sensing} with {Community} {Level} {Indicators}}. In
  \bibinfo{booktitle}{\emph{Proceedings of the 2018 {Designing} {Interactive}
  {Systems} {Conference}}} \emph{(\bibinfo{series}{{DIS} '18})}.
  \bibinfo{publisher}{Association for Computing Machinery},
  \bibinfo{address}{New York, NY, USA}, \bibinfo{pages}{1183--1192}.
\newblock
\showISBNx{978-1-4503-5198-0}
\urldef\tempurl%
\url{https://doi.org/10.1145/3196709.3196762}
\showDOI{\tempurl}


\bibitem[Dickinson et~al\mbox{.}(2018)]%
        {dickinson_inclusion_2018}
\bibfield{author}{\bibinfo{person}{Jessa Dickinson}, \bibinfo{person}{Sheena
  Erete}, \bibinfo{person}{Mark Diaz}, {and} \bibinfo{person}{Denise
  Linn~Riedl}.} \bibinfo{year}{2018}\natexlab{}.
\newblock \showarticletitle{Inclusion of {Underserved} {Residents} in {City}
  {Technology} {Planning}}. In \bibinfo{booktitle}{\emph{Extended {Abstracts}
  of the 2018 {CHI} {Conference} on {Human} {Factors} in {Computing}
  {Systems}}} \emph{(\bibinfo{series}{{CHI} {EA} '18})}.
  \bibinfo{publisher}{Association for Computing Machinery},
  \bibinfo{address}{New York, NY, USA}, \bibinfo{pages}{1--6}.
\newblock
\showISBNx{978-1-4503-5621-3}
\urldef\tempurl%
\url{https://doi.org/10.1145/3170427.3188583}
\showDOI{\tempurl}


\bibitem[Dow et~al\mbox{.}(2022)]%
        {dow_scaffolding_2022}
\bibfield{author}{\bibinfo{person}{Andy Dow}, \bibinfo{person}{Kyle Montague},
  \bibinfo{person}{Shauna Concannon}, {and} \bibinfo{person}{John Vines}.}
  \bibinfo{year}{2022}\natexlab{}.
\newblock \showarticletitle{Scaffolding {Young} {People}'s {Participation} in
  {Public} {Service} {Evaluation} through {Designing} a {Digital} {Feedback}
  {Process}}. In \bibinfo{booktitle}{\emph{Designing {Interactive} {Systems}
  {Conference}}} \emph{(\bibinfo{series}{{DIS} '22})}.
  \bibinfo{publisher}{Association for Computing Machinery},
  \bibinfo{address}{New York, NY, USA}, \bibinfo{pages}{319--334}.
\newblock
\showISBNx{978-1-4503-9358-4}
\urldef\tempurl%
\url{https://doi.org/10.1145/3532106.3533463}
\showDOI{\tempurl}


\bibitem[Gurstein(2011)]%
        {gurstein_open_2011}
\bibfield{author}{\bibinfo{person}{Michael~B. Gurstein}.}
  \bibinfo{year}{2011}\natexlab{}.
\newblock \showarticletitle{Open data: {Empowering} the empowered or effective
  data use for everyone?}
\newblock \bibinfo{journal}{\emph{First Monday}} (\bibinfo{date}{Jan.}
  \bibinfo{year}{2011}).
\newblock
\showISSN{1396-0466}
\urldef\tempurl%
\url{https://doi.org/10.5210/fm.v16i2.3316}
\showDOI{\tempurl}


\bibitem[Huron et~al\mbox{.}(2014)]%
        {huron_constructive_2014}
\bibfield{author}{\bibinfo{person}{Samuel Huron}, \bibinfo{person}{Sheelagh
  Carpendale}, \bibinfo{person}{Alice Thudt}, \bibinfo{person}{Anthony Tang},
  {and} \bibinfo{person}{Michael Mauerer}.} \bibinfo{year}{2014}\natexlab{}.
\newblock \showarticletitle{Constructive visualization}. In
  \bibinfo{booktitle}{\emph{Proceedings of the 2014 conference on {Designing}
  interactive systems}} \emph{(\bibinfo{series}{{DIS} '14})}.
  \bibinfo{publisher}{Association for Computing Machinery},
  \bibinfo{address}{New York, NY, USA}, \bibinfo{pages}{433--442}.
\newblock
\showISBNx{978-1-4503-2902-6}
\urldef\tempurl%
\url{https://doi.org/10.1145/2598510.2598566}
\showDOI{\tempurl}


\bibitem[Kambunga et~al\mbox{.}(2020)]%
        {kambunga_participatory_2020}
\bibfield{author}{\bibinfo{person}{Asnath~Paula Kambunga},
  \bibinfo{person}{Heike Winschiers-Theophilus}, {and}
  \bibinfo{person}{Rachel~Charlotte Smith}.} \bibinfo{year}{2020}\natexlab{}.
\newblock \showarticletitle{Participatory {Memory} {Making}: {Creating}
  {Postcolonial} {Dialogic} {Engagements} with {Namibian} {Youth}}. In
  \bibinfo{booktitle}{\emph{Proceedings of the 2020 {ACM} {Designing}
  {Interactive} {Systems} {Conference}}} \emph{(\bibinfo{series}{{DIS} '20})}.
  \bibinfo{publisher}{Association for Computing Machinery},
  \bibinfo{address}{New York, NY, USA}, \bibinfo{pages}{785--797}.
\newblock
\showISBNx{978-1-4503-6974-9}
\urldef\tempurl%
\url{https://doi.org/10.1145/3357236.3395441}
\showDOI{\tempurl}


\bibitem[Loukissas and Ntabathia(2021)]%
        {loukissas_open_2021}
\bibfield{author}{\bibinfo{person}{Yanni~Alexander Loukissas} {and}
  \bibinfo{person}{Jude~Mwenda Ntabathia}.} \bibinfo{year}{2021}\natexlab{}.
\newblock \showarticletitle{Open {Data} {Settings}: {A} {Conceptual}
  {Framework} {Explored} {Through} the {Map} {Room} {Project}}.
\newblock \bibinfo{journal}{\emph{Proceedings of the ACM on Human-Computer
  Interaction}} \bibinfo{volume}{5}, \bibinfo{number}{CSCW2}
  (\bibinfo{date}{Oct.} \bibinfo{year}{2021}), \bibinfo{pages}{357:1--357:24}.
\newblock
\urldef\tempurl%
\url{https://doi.org/10.1145/3479501}
\showDOI{\tempurl}


\bibitem[McAllister et~al\mbox{.}(2019)]%
        {mcallister_data_2019}
\bibfield{author}{\bibinfo{person}{Therese McAllister},
  \bibinfo{person}{Christopher Clavin}, \bibinfo{person}{Bruce Ellingwood},
  \bibinfo{person}{John van~de Lindt}, \bibinfo{person}{David Mizzen}, {and}
  \bibinfo{person}{Francis Lavelle}.} \bibinfo{year}{2019}\natexlab{}.
\newblock \showarticletitle{Data, information, and tools needed for community
  resilience planning and decision-making}. \bibinfo{publisher}{National
  Institute of Standards and Technology}, \bibinfo{address}{Gaithersburg, MD},
  \bibinfo{pages}{NIST SP 1240}.
\newblock
\urldef\tempurl%
\url{https://doi.org/10.6028/NIST.SP.1240}
\showDOI{\tempurl}
\newblock
\shownote{Report Number: NIST SP 1240}.


\bibitem[Puussaar et~al\mbox{.}(2018)]%
        {puussaar_making_2018}
\bibfield{author}{\bibinfo{person}{Aare Puussaar}, \bibinfo{person}{Ian~G.
  Johnson}, \bibinfo{person}{Kyle Montague}, \bibinfo{person}{Philip James},
  {and} \bibinfo{person}{Peter Wright}.} \bibinfo{year}{2018}\natexlab{}.
\newblock \showarticletitle{Making {Open} {Data} {Work} for {Civic}
  {Advocacy}}.
\newblock \bibinfo{journal}{\emph{Proceedings of the ACM on Human-Computer
  Interaction}} \bibinfo{volume}{2}, \bibinfo{number}{CSCW}
  (\bibinfo{date}{Nov.} \bibinfo{year}{2018}), \bibinfo{pages}{143:1--143:20}.
\newblock
\urldef\tempurl%
\url{https://doi.org/10.1145/3274412}
\showDOI{\tempurl}


\bibitem[Segel and Heer(2010)]%
        {segel_narrative_2010}
\bibfield{author}{\bibinfo{person}{Edward Segel} {and} \bibinfo{person}{Jeffrey
  Heer}.} \bibinfo{year}{2010}\natexlab{}.
\newblock \showarticletitle{Narrative {Visualization}: {Telling} {Stories} with
  {Data}}.
\newblock \bibinfo{journal}{\emph{IEEE Transactions on Visualization and
  Computer Graphics}} \bibinfo{volume}{16}, \bibinfo{number}{6}
  (\bibinfo{date}{Nov.} \bibinfo{year}{2010}), \bibinfo{pages}{1139--1148}.
\newblock
\showISSN{1941-0506}
\urldef\tempurl%
\url{https://doi.org/10.1109/TVCG.2010.179}
\showDOI{\tempurl}
\newblock
\shownote{Conference Name: IEEE Transactions on Visualization and Computer
  Graphics}.


\bibitem[Viegas et~al\mbox{.}(2007)]%
        {viegas_manyeyes_2007}
\bibfield{author}{\bibinfo{person}{Fernanda~B. Viegas}, \bibinfo{person}{Martin
  Wattenberg}, \bibinfo{person}{Frank van Ham}, \bibinfo{person}{Jesse Kriss},
  {and} \bibinfo{person}{Matt McKeon}.} \bibinfo{year}{2007}\natexlab{}.
\newblock \showarticletitle{{ManyEyes}: a site for visualization at internet
  scale}.
\newblock \bibinfo{journal}{\emph{IEEE transactions on visualization and
  computer graphics}} \bibinfo{volume}{13}, \bibinfo{number}{6}
  (\bibinfo{year}{2007}), \bibinfo{pages}{1121--1128}.
\newblock
\showISSN{1077-2626}
\urldef\tempurl%
\url{https://doi.org/10.1109/TVCG.2007.70577}
\showDOI{\tempurl}


\bibitem[Vlachokyriakos et~al\mbox{.}(2016)]%
        {vlachokyriakos_digital_2016}
\bibfield{author}{\bibinfo{person}{Vasillis Vlachokyriakos},
  \bibinfo{person}{Clara Crivellaro}, \bibinfo{person}{Christopher~A.
  Le~Dantec}, \bibinfo{person}{Eric Gordon}, \bibinfo{person}{Pete Wright},
  {and} \bibinfo{person}{Patrick Olivier}.} \bibinfo{year}{2016}\natexlab{}.
\newblock \showarticletitle{Digital {Civics}: {Citizen} {Empowerment} {With}
  and {Through} {Technology}}. In \bibinfo{booktitle}{\emph{Proceedings of the
  2016 {CHI} {Conference} {Extended} {Abstracts} on {Human} {Factors} in
  {Computing} {Systems}}} \emph{(\bibinfo{series}{{CHI} {EA} '16})}.
  \bibinfo{publisher}{Association for Computing Machinery},
  \bibinfo{address}{New York, NY, USA}, \bibinfo{pages}{1096--1099}.
\newblock
\showISBNx{978-1-4503-4082-3}
\urldef\tempurl%
\url{https://doi.org/10.1145/2851581.2886436}
\showDOI{\tempurl}


\end{thebibliography}

\end{document}